# Fast Fabrication of Sub-200-nm Nanogrooves using Liquid Crystal Material

*Dae Seok Kim, Yun Jeong Cha, Min-Jun Gim and Dong Ki Yoon\**

Graduate School of Nanoscience and Technology, KAIST, Daejeon 305-701, Republic of Korea

*Email : nandk@kaist.ac.kr

**ABSTRACT**

Self-assembly of soft materials attracts keen interest for patterning applications owing to its ease and spontaneous behavior. We report the fabrication of nanogrooves using sublimation and recondensation of liquid crystal (LC) materials. First, well-aligned smectic LC structures are obtained on the micron-scale topographic patterns of the microchannel; then the sublimation and recondensation process directly produces nanogrooves having sub-200-nm scale. The entire process can be completed in less than 30 min. After it is replicated using an ultraviolet-curable polymer, our platform can be used as an alignment layer to control other guest LC materials.

# 1. INTRODUCTION

Lithographic tools using soft materials such as colloidal particles, block copolymers (BCPs), and liquid crystals (LCs) have been extensively studied to overcome the limits of current photolithography-based fabrication methods.[1–9] For this reason, directed self-assembly using soft materials has been gaining importance in "The International Technology Roadmap for Semiconductors,"[10] although there are still many obstacles to their practical application.[11,12] Among the various types of soft materials used in this fabrication application, the BCP is the most intensively studied and well-defined material; its typical periodicity is less than 50 nm owing to the characteristic microphase separation between the blocks.[4–7,13,14] The periodic nanostructures of BCPs facilitate a variety of applications including template patterns and many other applications.[15–19] To extend its application to optical response at visible or infrared wavelengths,[20,21] larger-scale patterns (more than 100 nm) are required, though they were proven to be difficult to obtain with this strategy owing to the poor translational ordering.[22–24]

Here, we developed highly periodic nanogrooves with a sub-200-nm feature size over large areas using a smectic LC material that can sublime and recondense to form secondary structures.[25,26] Unlike previously reported self-assembling materials, our platform takes advantage of the ability to achieve fast fabrication of sub-200-nm patterns in less than 30 min and obtain robust structures to mold other materials. To realize this goal, first, long-range (centimeter-scale) ordered smectic LC domains were prepared using micron-scale topographic patterns made of silicon. Then, a mild thermal sublimation and recondensation process was performed to generate nanogrooves that could be transferred to an ultraviolet (UV)-curable polymer by microcontact printing. Our platform was further employed to use the polymer nanogrooves as an alignment layer for guest nematic LC molecules. This simple and rapid process offers the potential for nanometer-scale surface patterning by self-assembly of LCs and can be a cornerstone for more sophisticated lithographic applications based on smectic LC materials and other soft materials.

## 2. RESULT AND DISCUSSION

**Fabrication of Nanogrooves.** Figure 1 illustrates the process used to prepare well-ordered nanogrooves. A fluorinated LC material, here **Y002** that can sublime at its smectic A (SmA) LC phase was used in this process (Figure 1a). A sandwich cell with a silicon microchannel on the bottom and a glass cover slip was prepared, and all the substrates were spin-coated with polyethyleneimine (PEI) to induce planar arrangement of LC molecules, which was not rubbed (Figure 1b). **Y002** in the isotropic melt state (~200 °C) was injected between the substrates by capillary action. When it was cooled at 0.5 °C/min from the isotropic state to the high-temperature SmA LC phase (~192 °C), LC molecules were planar- and parallel-aligned on the substrate along the channels (Figure 1c). Then, the sample was cooled to room temperature (~25 °C) at a rapid rate of 30 °C/min to quench the smectic LC phase.[8,25] The cover slip was removed, and the sample was thermally annealed at 120 °C for 5 min to generate nanogrooves on top of the LC film (Figure 1d,e). Because a low-molecular-weight LC materials such as the one used here is susceptible to heat,[26] and thus the fabrication of nanogrooves is very fast, we found that nanogrooves were generated over a large area in less than 30 min. To extend their application to soft lithography, UV-curable polymer was used to mold the nanogrooves (Figure 1f,g).

To realize this platform, a preliminary test to observe the nanogrooves was conducted (Figure 2). A sample prepared between untreated glass substrates, which can induce planar anchoring of the LC molecules, shows typical fan-shaped textures in a polarized optical microscopy (POM) image (Figure 2a). After the cover slip was removed, scanning electron microscopy (SEM) was used to observe the detailed surface structure, revealing a very flat texture overall, with some cracks that were generated during quenching of the sample (Figure 2c). An enlarged atomic force microscopy (AFM) image shows a root mean square surface roughness (Rq) of ~5 nm and plain textures, as shown in the depth profiles (Figure 2e,g). In contrast, the slightly sublimed and recondensed sample treated at 120 °C for 5 min shows tiny dark lines in a POM image (Figure 2b). These lines are directly revealed in a SEM image (Figure 2d), which shows that linear structures (referred to as nanogrooves hereafter) were generated, although their orientation is poor, as they follow the random planar textures of the LC film. Without any type of etching process, the patterns themselves have topographical nanogroove shapes, as directly determined by a cross-sectional SEM image (inset of Figure 2d). The corresponding AFM image shows nanogrooves that have Rq ~ 105 nm, and the depth profile also shows how the nanogrooves appear on the surface (Figure 2f,h). As shown in Figure 2h, the nanogrooves exhibit roughly periodic features, with a width (w) of ~200 nm and a depth (d) of ~30 nm. The nanogrooves with free surface were preserved until the isotropic phase, ~195 °C on heating (Figure S1). If the temperature reached isotropic phase and then cooled down to the SmA phase again, the linear arrays would be replaced by toric focal conic domains (TFCDs) as previously reported.[8,9]

This shows a clear evidence for the reorganization of the LC/air interface sketched in Figures 2c and 2d, where the surface structures transform into the multi-Burgers vector edge dislocation structures, leading to form array of linear nanogrooves with hemi-cylindrical shape (inset of Figure 2d). The energy/area, W, of the surface for the planar structure is $W = \sigma_\parallel$, the surface tension with the layers normal to air interface (Figure 2c), and for the nanogrooves, it is $W = \pi\sigma_\perp/2$, where $\sigma_\perp$ is the surface tension with the layers parallel to the air interface (Figure 2d) and we have excluded the relatively small curvature elastic energy.[27] The strong preference for the nanogrooves shows that $\sigma_\perp$ must be much smaller than $2\sigma_\parallel/\pi$, thus, $\sigma_\perp$ is the operative surface tension of the interface.[25] Thus, the frustration inherent in the competing orientation preferences at the LC/air interface is resolved by the formation of nanogrooves to make the vertical layers end up parallel to the LC/air interface during the sublimation and recondensation process in the SmA phase (Figure 2d).

Periodic curvature on the order of a few hundred nanometers has not been observed in rod-like smectic LC materials to date, because the elastic energy density allows only the formation of bulk curvature in microscale.[28,29] However, as recently reported,[25,26] mild thermal treatment facilitates anisotropic reorientation of LC material on the nanoscale through sublimation and recondensation with a very small loss (less than 1 wt%) of LC material on the free surface of the SmA film, resulting in the formation of nanogrooves. At this point, we recognized the importance of the well-ordered smectic layers to realize our goal.

Controlling the orientation of thermotropic smectic phases that have no nematic phase in a higher-temperature region is generally not as easy as controlling that of the simple nematic phase; thus, many types of new tools have been proposed to align smectics, including topographic confinement, epitaxial growth, and chemical patterning.[28,30–31] In particular, confined geometries using topographic confinement were proven to be very useful for controlling the defect structures of smectic phases.[8,28,30–33] Here, the sandwich cell composed of a silicon microchannel and glass cover slip was used to control the molecular director, n, which is called the layer normal vector. As in the report by Clark and coworkers on smectics on trenched substrates,[30] almost a single domain of the SmA phase was also obtained in our experiment. Two POM images observed using different polarizer angles show a dramatic change in the light transmission (Figure 3a,b), where all the molecules are aligned parallel to the channel direction, as shown in the schematic image in the inset of Figure 3b. The upper-right insets in Figure 3 show the depolarized images and the channel direction. During formation of the single SmA domain of the LC film, two important factors were considered to control the alignment of LC molecules. First, the microchannels provide a confinement effect inducing anisotropic orientation of the batonnet-like SmA nucleation in microscale, so molecules are likely to be aligned parallel to the channel direction, which can be understood by considering the elastic splay modulus, because the bend modulus is prohibited in the SmA phase by divergence of the elastic energy constant, $K_3$.[30,34] Thus, layers of SmA batonnets are preferentially grown perpendicular to the channel direction. Second, we also observed that the slow cooling rate of 0.5 °C/min causes nucleation and growth of SmA batonnets from the edges of the substrate bordering the air interface, producing a thermal gradient from the edge to the center of the substrate. Thus, the relatively slow cooling rate may suppress growth of SmA batonnets in random areas of the substrate but foster continuous growth from the edge into a single domain.

The sublimation and recondensation process was applied to this well-aligned smectic film under experimental conditions identical to those in the previous test (Figure 2b). As expected, long-range-ordered nanogrooves were fabricated (Figure 4a), the featured pattern size of which was around 200 nm. The greatest difference between our system and those in previous LC-based works is its high robustness.[29,35] Consequently, we could replicate our pattern with other soft materials. Here, the inverse replica of the nanogrooves was obtained using the UV-curable prepolymer Norland Optical Adhesive 63 (NOA63). Capillary forces drive the prepolymer into the nanogrooves and cause it to conform to the surface topography. The prepolymer is then cured to form a solid by irradiating it with UV light (365 nm), resulting in a negative copy of the nanogrooves. The pattern of the NOA63 polymer stamp was peeled from the nanogrooves master, showing that highly dense nanogrooves were successfully transferred from the smectic LC master to the NOA63 polymer stamp, as shown in a SEM image (Figure 4b). The AFM image and depth profile show the sharp features of the molded patterns, which exhibit good preservation of the inverse geometry of the original LC nanogrooves (Figure 4c,d).

**Application of Nanogrooves.** On the basis of the results, we tried to use our platform in an LC alignment layer application (Figure 5). A sandwich cell was fabricated in which the NOA63 replica of the nanogroove pattern was used as the bottom substrate, and the top cover slip was coated with a planar-aligned polyimide layer without rubbing. The cell thickness was fixed at 3 μm by silica ball spacers. Then the commercially available rod-like nematic LC 4-cyano-4′-n-pentylbiphenyl (5CB) was injected into the cell by capillary force at isotropic temperature (40 °C). The orientation order of 5CB molecules, $\mathbf{n}_{5CB}$, was aligned parallel to the nanogrooves in the nematic phase, and the entire area of the LC film shows a homogeneous texture (Figure 5a,b). For comparison with the controlled experiment, Figure 5c shows nematic textures prepared between untreated glass substrates, in which randomly grown schlieren textures appear. Our nanogroove LC cell shows dramatic birefringence changes when the sample is rotated between the polarizer and analyzer. When $\mathbf{n}_{5CB}$ is parallel to either the polarizer or the analyzer, the nematic domain exhibits good extinction (Figure 5a), whereas it becomes bright when $\mathbf{n}_{5CB}$ is tilted at 45° with respect to the crossed polarizers (Figure 5b).

To evaluate the aligned nematic phase in the nanogroove–cover slip cell, optical analysis was conducted as a function of the rotation angle ($\theta$) between $\mathbf{n}_{5CB}$ and the polarizer (Figure 5d). The LC domain shows a highly symmetric sinusoidal intensity profile as a function of $\theta$. The highest transmittance appears at $\theta = 45$ and 135°. This can be simply explained by the light-guiding effect of well-aligned LC molecules, as shown in a conventional LC cell made by the mechanical rubbing method.[36] Further, the dark state appears when the molecular directors are parallel or perpendicular to the polarizers ($\theta$ = 0, 90, or 180°). In contrast, the random nematic domains show almost constant intensity over the entire θ range, corresponding to the averaged intensity of multiple nematic domains (Figure 5c,d).

The pretilt angle and azimuthal anchoring energy were also estimated to show that our nanogrooves are comparable with the conventional alignment layer made by rubbed polyimide (PI). Transmittance tendency was plotted as a function of rotation angle of the LC cell based on UV-treated NOA63 substrate replicated by nanogrooves along with a polar direction under crossed polarizers (Figure S2). The pretilt angle of LC molecules from the nanogrooves was about 3.7°, which is comparable to that of the rubbed PI based LC cell (2−3°).[37] Then, the azimuthal anchoring energy $W_a$ of the sinusoidal grooves by the followed equation;[38] $W_a = \frac{2\pi^3 K A^2}{\lambda^3}$. In this expression, $A$ is the groove depth, $\lambda$ is the groove pitch, $K$ is approximated elastic constant. For nematic LC, 5CB ($K \sim 10^{-11}$N), the corresponding anchoring energy, is $\sim 0.7 \times 10^{-5}$ J/m$^2$ which is comparable to the value obtained in the PI alignment layer, $\sim 10^{-4}$ J/m$^2$.[39]

## 3. CONCLUSION

In summary, we first report that the smectic LC phase can be used to make nanogrooves by a combination of micro-confinement and a thermal sublimation and recondensation process. Submicron fabrication is usually difficult and expensive, but in this work it was realized using one type of LC material and a simple Si microchannel, so it costs much less than conventional methods. Indeed, it takes only 30 min or less, which is useful for application in a parallel process of fabricating nanostructures with other fabrication tools, including conventional ones. We believe that the resulting platform can be applied broadly in patterning applications.

## 4. EXPERIMENTAL SECTION

*Materials:* The LC material **Y002** was prepared as previously reported.[8] Phase transitions of the samples were characterized by differential scanning calorimetry (Q1000 V9.9 Build 303).

*Sample Preparation (Fabrication of nanogrooves)*: A Si microchannel with a featured width of 5 µm and depth of 5 µm was prepared and spin-coated with PEI (Aldrich; MW: 60,000) for planar anchoring and covered by a cover glass, which also provides planar anchoring, with a gap of ~3 µm (Figure 1b). Then, a crystalline powder of **Y002** was heated to isotropic temperature (200 ℃) on a heating stage (LINKAM LTS420) equipped with a temperature controller (LINKAM TMS94) to fill the sandwich cell by capillary action and cooled to the high-temperature SmA region (~192 ℃) at 0.5 ℃/min, during which SmA batonnets grew continuously and gradually to form a single domain. After the phase transition into SmA was complete, the sample was quenched to 25 ℃ and the cover glass was removed; then the sample was thermally annealed at 120 ℃/min for 5 min to generate the nanogrooves. The UV-curable polymer NOA63 (Norland Products, Inc.), was poured onto the LC nanogrooves and then cured by UV irradiation using a UV lamp (Vilber Lourmat, power: 8 W) with a 365 nm tube for 2 min. Then, the replicated pattern was carefully removed at 70 ℃.

*Characterization***:** The optical textures of the prepared LC film, LC nanogrooves, and 5CB on the nanogrooves were measured using POM (Nikon LV100POL). The LC nanogrooves and their inverse mold of NOA63 polymer were observed by field-emission SEM (FE-SEM; Hitachi, S-4800) at 7 kV and 7 µA after a 5-nm-thick Pt coating was applied. The surface structures of the LC films and replica were investigated using AFM (Bruker, Multimode-8) with a 100 µm$^2$ scanner in tapping mode under ambient conditions; for this purpose, an antimony-doped silicon cantilever with a spring constant of 20–80 N/m and frequency ($f_0$) of 300 kHz was used.

AUTHOR INFORMATION

**Corresponding Author**

*E-mail: nandk@kaist.ac.kr

**Author Contributions**

D.S.K. and D.K.Y. designed the research; D.S.K., Y.J.C. and M.J.G. performed the research; D.S.K. and D.K.Y. analyzed the data; and D.S.K. and D.K.Y. wrote the manuscript. All authors have given approval to the final version of the manuscript.

**Notes**

The authors declare no competing financial interest.

ACKNOWLEDGMENT

This work was supported by a grant from the National Research Foundation (NRF) and funded by the Korean Government (MSIP) (2014M3C1A3052567 and 2015R1A1A1A05000986).

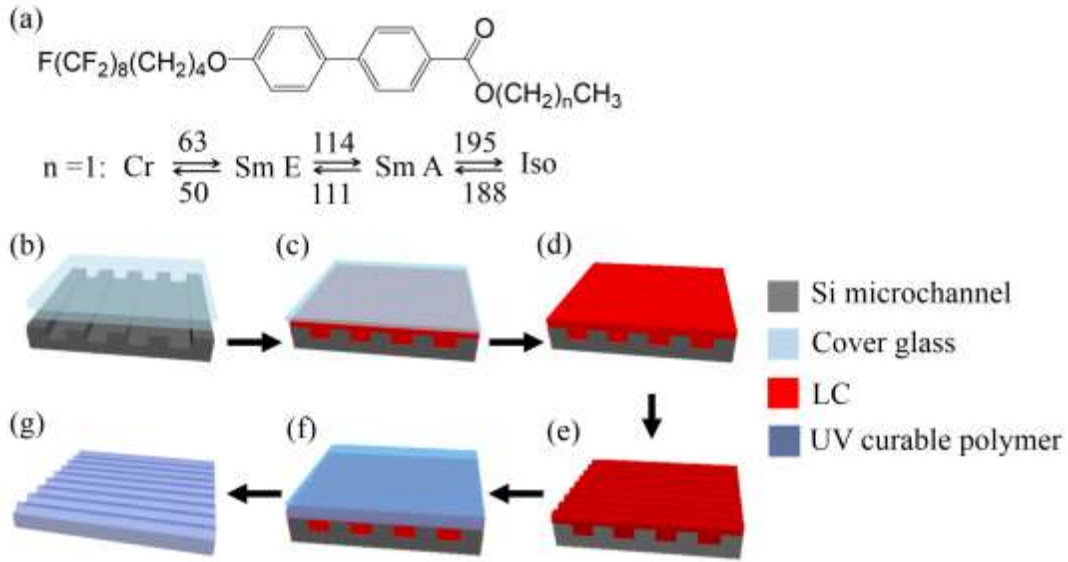

**Figure 1.** LC material and experimental schemes used to fabricate nanogrooves. (a) Chemical structure and phase diagram of the semifluorinated LC molecule **Y002**. (b) A sandwich cell made of a microchannel and glass cover slip was prepared. (c) The cell was filled with **Y002** in the isotropic phase by capillary action, and **Y002** was cooled to the SmA phase. (d) The cover glass was removed to expose the upper surface of the LC film to air. (e) Through mild thermal treatment, uniform nanogrooves were generated. (f) UV-curable prepolymer was put onto the nanogrooves and cured by UV irradiation. (g) After completely hardening, the cured polymer substrate was carefully removed to show the inverse patterns of the LC nanogrooves.

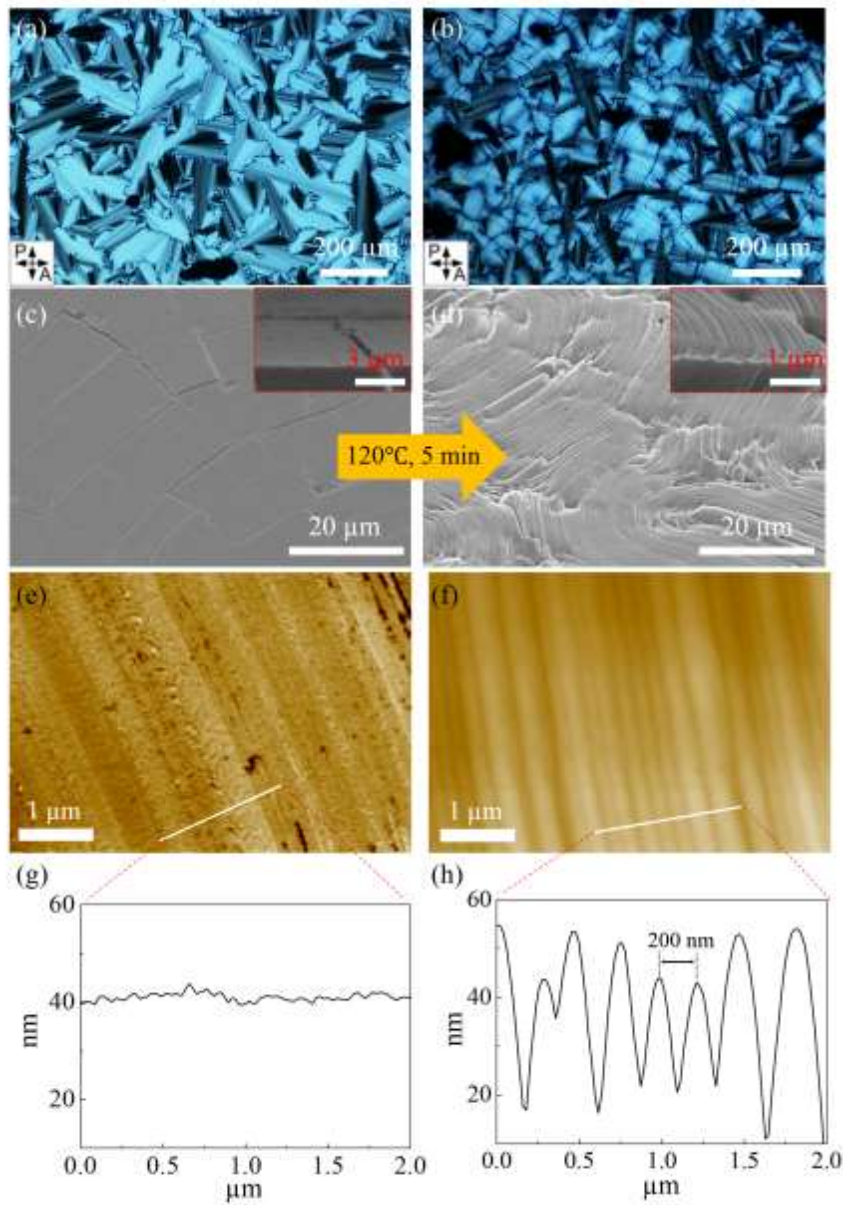

**Figure 2.** Characterization of the original SmA LC film and nanogrooves. (a,b) POM images of the original and thermally annealed SmA LC film, respectively. (c,d) SEM images corresponding to the SmA structures before and after thermal annealing, respectively; insets show magnified views. (e,f) AFM images of the original LC film and LC nanogrooves, respectively. Root mean square surface roughness ($R_q$) of the samples changed from ~5 nm to ~106 nm after thermal annealing. (g,h) Corresponding depth profiles for the original LC film and nanogroove film.

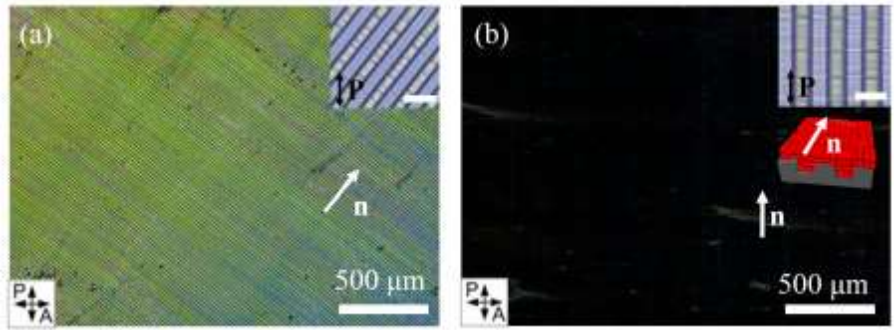

**Figure 3.** POM images of the film with a single large domain of SmA LC generated in the microchannel–cover slip cell. (a,b) POM images of the single-domain lamellar LC film at +45° and 0° with respect to the analyzer, respectively. White arrows indicate the molecular director, **n**. Inset in each figure is a depolarized image showing the channel direction for each state, and the schematic sketch in (b) shows the molecular arrangement in the cell. All scale bars in insets are 10 µm.

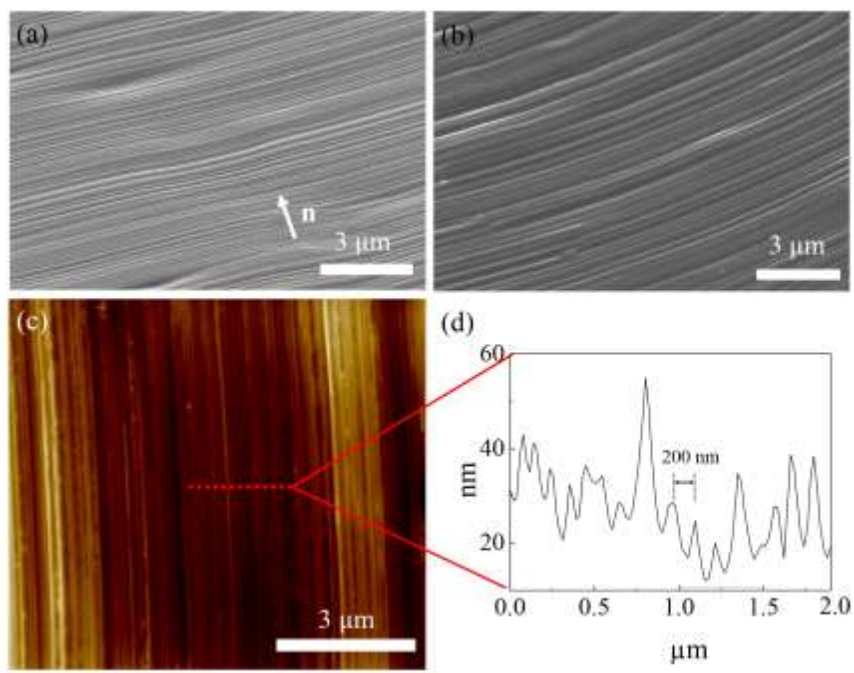

**Figure 4.** LC nanogrooves and their replica. SEM images of (a) LC nanogrooves and (b) NOA63 replica. (c) AFM image of the molded-in NOA63 and (d) its depth profile, showing inverse replicated patterns of the LC nanogrooves.

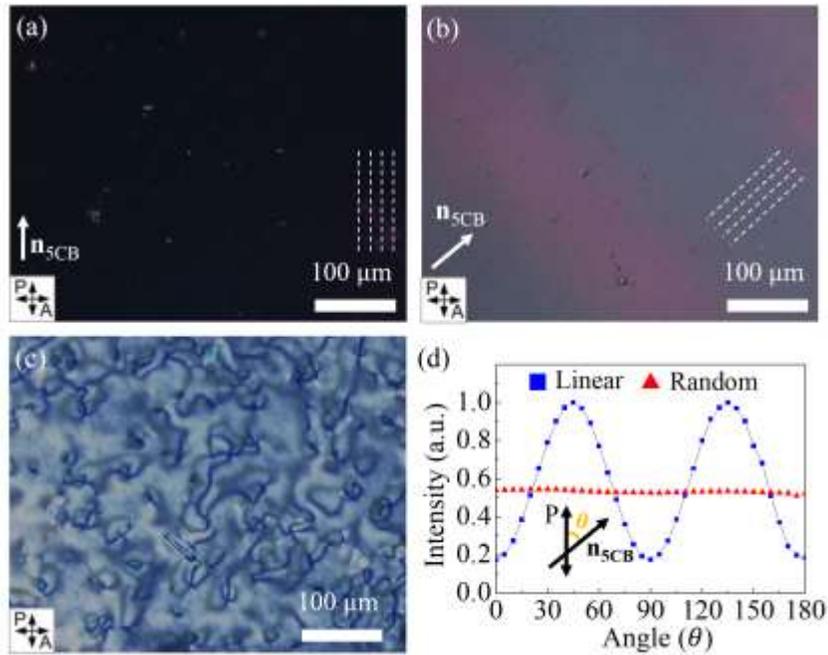

**Figure 5.** Nematic LC cell consisting of the NOA63 replica and a glass cover slip. (a) When the long axis of the nanogrooves is parallel to the analyzer, a POM image shows good extinction, meaning that **n**$_{5CB}$ is well-aligned along the nanogroove direction. (b) When **n**$_{5CB}$ is placed at 45° to the analyzer, a POM image shows the highest birefringence intensity. (c) POM image of nematic LC prepared between untreated cover glasses shows typical schlieren textures. (d) Birefringence intensity profile as a function of rotation angle ($\theta$) between **n**$_{5CB}$ and polarizer (P).

# Supporting Information

# Fast Fabrication of Sub-200-nm Nanogrooves using Liquid Crystal Material

Dae Seok Kim, Yun Jeong Cha, Min-Jun Gim and Dong Ki Yoon*

Graduate School of Nanoscience and Technology, KAIST, Daejeon 305-701, Republic of Korea

* Corresponding Authors Email: nandk@kaist.ac.kr

**Figure S1.**

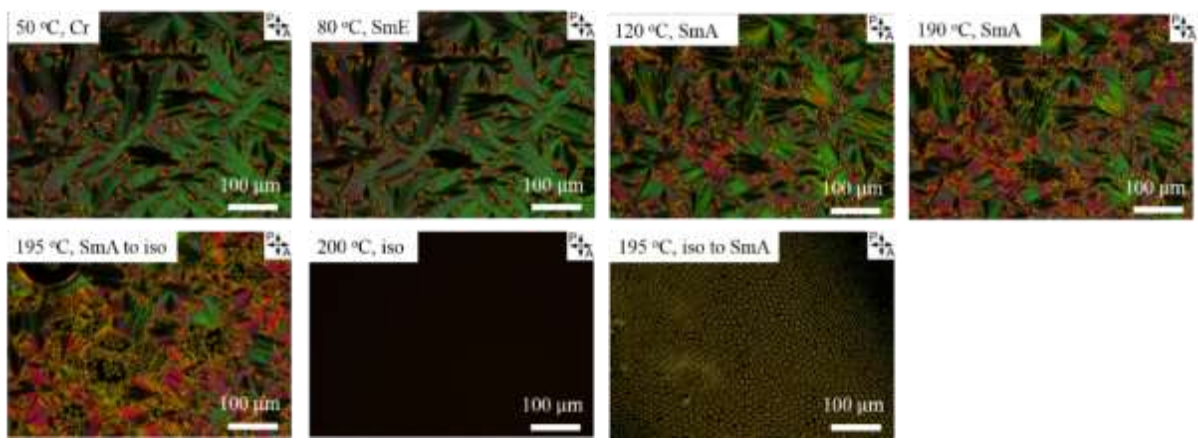

**Figure S1.** The POM images of planar LC film with free surface during heating from crystal phase to isotropic and then cooling down into SmA phase again.

**Figure S2**.

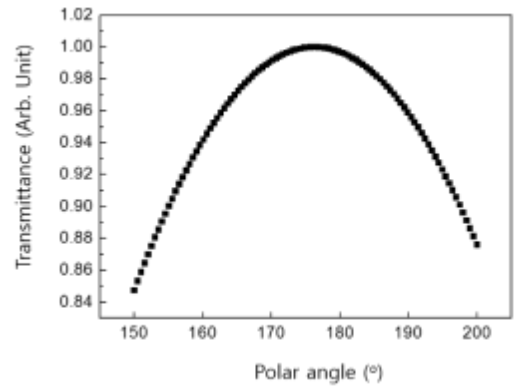

**Figure S2.** Transmittance tendency as a function of rotation angle of the LC cell based on UV-treated NOA63 substrate patterned by nanogrooves along with polar direction under crossed polarizers.